\documentclass[a4paper]{jpconf}
\usepackage[pdftex]{graphicx}
\usepackage{dcolumn}
\usepackage{bm}
\usepackage{color}
\usepackage{pifont}
\usepackage{hyperref}
\hypersetup{
colorlinks = true,
urlcolor   = blue,
linkcolor  = blue,
citecolor  = blue
}
\begin{document}
\title{Magnetic nature of wolframite MgReO$_4$}

\author{Elisabetta~Nocerino$^{1,*}$, Ola~K.~Forslund$^2$, Chennan~Wang$^3$, Hiroya~Sakurai$^4$, Frank~Elson$^1$, Rasmus~Palm$^1$, Ugne~Miniotaite$^1$, Yuqing Ge$^2$, Yasmine~Sassa$^2$, Jun~Sugiyama$^5$, Martin~M{\aa}nsson$^{1,+}$}
\address{$^1$Department of Applied Physics, KTH Royal Institute of Technology, SE-106 91, Stockholm, Sweden}
\address{$2$Department of Physics, Chalmers University of Technology, SE-412 96 G$\ddot{o}$teborg, Sweden}
\address{$^3$Laboratory for Muon Spin Spectroscopy, Paul Scherrer Institute, 5232 Villigen PSI, Switzerland}
\address{$^4$National Institute for Materials Science, Namiki, Tsukuba, Ibaraki 305-0044, Japan}
\address{$^5$Neutron Science and Technology Center, Comprehensive Research Organization for Science and Society (CROSS), Tokai, Ibaraki 319-1106, Japan}

\ead{$^{*}$nocerino@kth.se, $^{+}$condmat@kth.se}

\begin{abstract}
Rhenium oxides belonging to the family $A$ReO$_4$ where $A$ is a metal cation, exhibit interesting electronic and magnetic properties. In this study we have utilized the muon spin rotation/relaxation ($\mu^+$SR) technique to study the magnetic properties of the MgReO$_4$ compound. To the best of our knowledge, this is the first investigation reported on this interesting material, that is stabilized in a wolframite crystal structure using a special high-pressure synthesis technique. Bulk magnetic studies show the onset of an antiferromagnetic (AF) long range order, or a possible singlet spin state at $T_{\rm C1}\approx90$~K, with a subtle second high-temperature transition at $T_{\rm C2}\approx280$~K. Both transitions are also confirmed by heat capacity ($C_p$) measurements. From our $\mu^+$SR measurements, it is clear that the sample enters an AF order below $T_{\rm C1}=T_{\rm N}\approx85$~K. We find no evidence of magnetic signal above $T_{\rm N}$, which indicates that $T_{\rm C2}$ is likely linked to a structural transition. Further, via sensitive zero field (ZF) $\mu^+$SR measurements we find evidence of a spin reorientation at $T_{\rm Cant}\approx65$~K. This points towards a transition from a collinear AF into a canted AF order at low temperature, which is proposed to be driven by competing magnetic interactions.

\end{abstract}

\section{Introduction}
\begin{figure}[h]
  \begin{center}
    \includegraphics[scale=0.99]{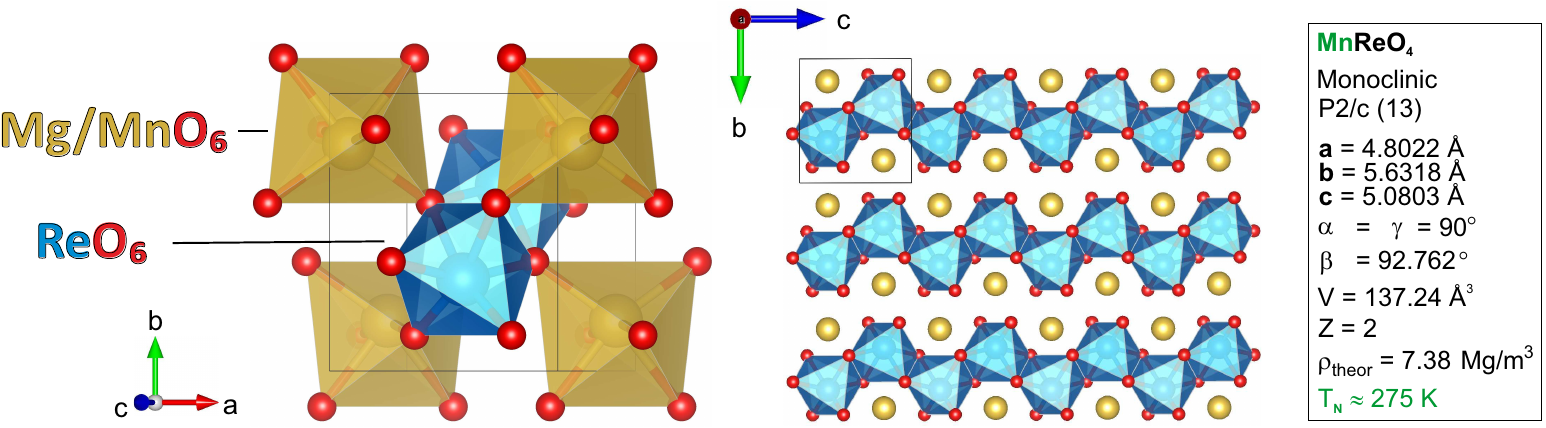}
  \end{center}
  \caption{The wolframite crystal structure of MgReO$_4$ and room temperature crystallographic lattice parameters of the closely related MnReO$_4$ compound, which display an antiferromagnetic transition at $T_{\rm N}\approx275$~K \cite{bramnik2005preparation}.}
  \label{struc}
\end{figure}

Rhenium oxides \cite{Jana_2019,Konpap,Forslund_2020} of the type $A$ReO$_4$, where $A$ is usually a first-row transition metal and Re is found in a oxidation state lower than 7+, are not very common and the scientific literature presents only few examples of such materials \cite{sleight1975new,baur1992coreo4,urushihara2021structural}. Compounds with Re in an oxidation state 6+ are expected to manifest unusual physical properties due to the strong spin-orbit coupling of the Re 5$d^1$ electrons \cite{chen2010exotic,bramnik2001preparation,bramnik2003preparation}, which could introduce quantum fluctuations and lead to multipolar magnetic ordering. The very first material of this kind, MnReO$_4$, was synthesized in the 1970's together with MgReO$_4$ and ZnReO$_4$ \cite{sleight1975new}. When first reported, it was suggested that the three samples possessed a wolframite structure, based on the refinement of the unit cell and in qualitative agreement with analogous tungstates and molybdates. While the exact crystal structure, with the refinement of atomic positions, has not been reported yet for MgReO$_4$ and ZnReO$_4$, a recent x-ray diffraction study confirmed the wolframite structure for MnReO$_4$ \cite{bramnik2005preparation}. The latter material also manifested an anisotropic electrical resistivity, which is an attractive feature for potential applications in the development of electrical devices \cite{patent}.
The wolframite-type structure can be described as a distorted hexagonal close packing of O atoms with each of the $A$ and Re atoms occupying one-fourth of the octahedral voids. The distorted octahedra, containing cations of only one type, form zig-zag chains extended along the $c$-axis. Octahedra of the same type are joined by edges, octahedra of different type are joined by corners. The wolframite structure of AReO$_4$ is displayed in Fig.~\ref{struc}.

From a comparison among bond-lengths in different Re and Mn oxides, it was suggested that the electronic configuration in MnReO$_4$ would be Mn$^{2+}$Re$^{6+}$O$_4^{2-}$ with a half filled Mn-3$d^5$ and Re-5$d^1$ system \cite{bramnik2005preparation} as in Ba$_{2-x}$Sr$_x$MnReO$_6$ \cite{popov2003large}. For MgReO$_4$ and ZnReO$_4$, the definite oxidation state 2+ for both Mg and Zn ions simply leads to a Re-5$d^1$ system with a delocalized nature of the valence electron \cite{sleight1975new}.

These materials represent a potentially interesting study case and, even though they were already known for decades, their physical properties have not been extensively studied and, for MgReO$_4$ and ZnReO$_4$, no further report beyond the original synthesis paper have been published. In this work we present the very first investigation of the magnetic nature of the wolframite insulator MgReO$_4$ carried out by both bulk methods as well as muon spin rotation/relaxation technique ($\mu^+$SR). The occurrence of a static long range magnetic ordering was clearly observed in the muon spectra and the order parameter of the magnetic phase transition was determined. Finally, by the unique sensitivity of the zero field $\mu^+$SR protocol we are able to reveal the onset of a possible spin canting inside the antiferromagnetic phase as temperature is further decreased.

\section{Experimental Methods}

A high pressure synthesis technique ($p=6$~GPa at 1300~C$^{\circ}$ for 1 hr) was adopted at the National Institute for Materials Science (NIMS) in Japan to produce high-quality powder samples of MgReO$_4$. The $\mu^+$SR measurements were performed at the General Purpose Surface-Muon Instrument (GPS) of the Swiss Muon Source (S$\mu$S) at the Paul Scherrer Institute (PSI). Approximately 200~mg of powder was loaded into an envelope made of 30 $\mu$m thick aluminum-coated Mylar tape, which was in turn sealed with thin kapton tape to create a hermetic seal from air and moisture. The entire sample handling was conducted inside a helium glove box to avoid any sample degradation. The envelope was suspended in the muon beam using a fork-type sample holder and loaded in a flow cryostat (temperature range: 1.5 K to 300 K). Zero field (ZF) and weak transfer field (wTF) protocols were used to systematically acquire $\mu^+$SR time spectra for a set of different temperatures. From the ZF experiment it was possible to investigate the sample's internal magnetic environment and outline the magnetic order parameter, while the wTF setup allowed to estimate the magnetic transition temperature and the magnetic volume fractions in the sample through the application of an external magnetic field of 50 G, with its direction oriented transversally with respect to the initial direction of the muon spin polarization. More details about the $\mu^+$SR technique and experimental setup can be found elsewhere \cite{blundell1999spin,Blundell_2021}.

\begin{figure}[h]
  \begin{center}
    \includegraphics[scale=0.6]{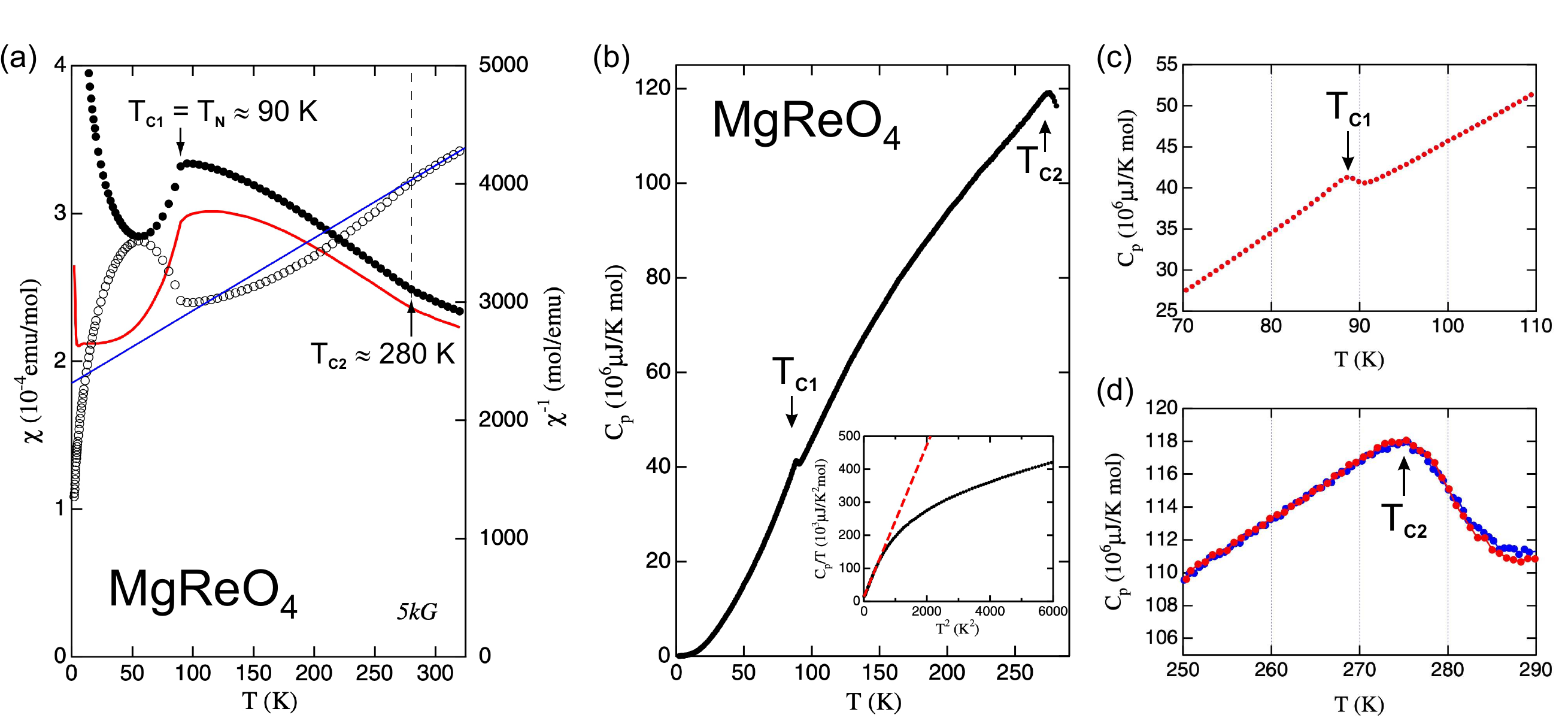}
  \end{center}
  \caption{(a) Temperature dependencies of $\chi$ (solid circle) and inverse $\chi$ (open circle) measured with a magnetic field of 5 kG. A clear antiferromagnetic transition is revealed at $T_{\rm C1}=T_{\rm N}\approx90$~K. The red line shows the corrected $\chi$(T) curve obtained by subtraction of the magnetic impurities contribution. The blue line highlights the deviation of the susceptibility from linearity below $T_{\rm C2}$. (b) Temperature dependence of the heat capacity (C$_P$) in MgReO$_4$ in the full temperature range measured on heating and cooling. The inset shows the fitting curve for the estimation of the Debye temperature. The magnification of C$_P$(T) around the transition temperatures (c) $T_{\rm C1}=T_{\rm N}$ and (d) $T_{\rm C2}$, show that there is no hysteresis between the data measured on heating (red circles) and cooling (blue circles).}
  \label{bulk}
\end{figure}

Initial bulk characterizations were carried out on the sample prior to the $\mu^+$SR experiment. Figure~\ref{bulk}(a) shows the temperature dependence of magnetic susceptibility ($\chi$) and inverse $\chi$ of MgReO$_4$, measured under a magnetic field of 5 kG, together with a corrected curve, which accounts for the presence of magnetic impurities. As the temperature decreases from 320 K, the corrected $\chi$ (red line) increases with a change in its slope (d$\chi$/dT) at T$_{\rm C2}$ = 280 K, and reaches a local maximum at $T_{\rm C1}$ = 90 K. Then, $\chi$ abruptly decreases below T$_{\rm C1}$ and levels off to a constant value below about 20 K. This behavior indicates that MgReO$_4$ enters into either an antiferromagnetically (AF) ordered state or a spin singlet state below $T_{\rm C1}$, while the nature of the magnetic state in the temperature range between $T_{\rm C1}$ and $T_{\rm C2}$ is not clear. Heat capacity ($C_p$) measurements [Fig.~\ref{bulk}(b-d)] also show the presence of two transitions at $T_{\rm C1}=90$~K and at $T_{\rm C2}=275$~K with no temperature hysteresis. The latter is a possible indication for a second order-type phase transition. The heat capacity curve also allowed to estimate the sample's Debye temperature as $\Theta_{\rm D}=120$~K.

\section{Results and Discussion}

\begin{figure}[h]
  \begin{center}
    \includegraphics[scale=0.95]{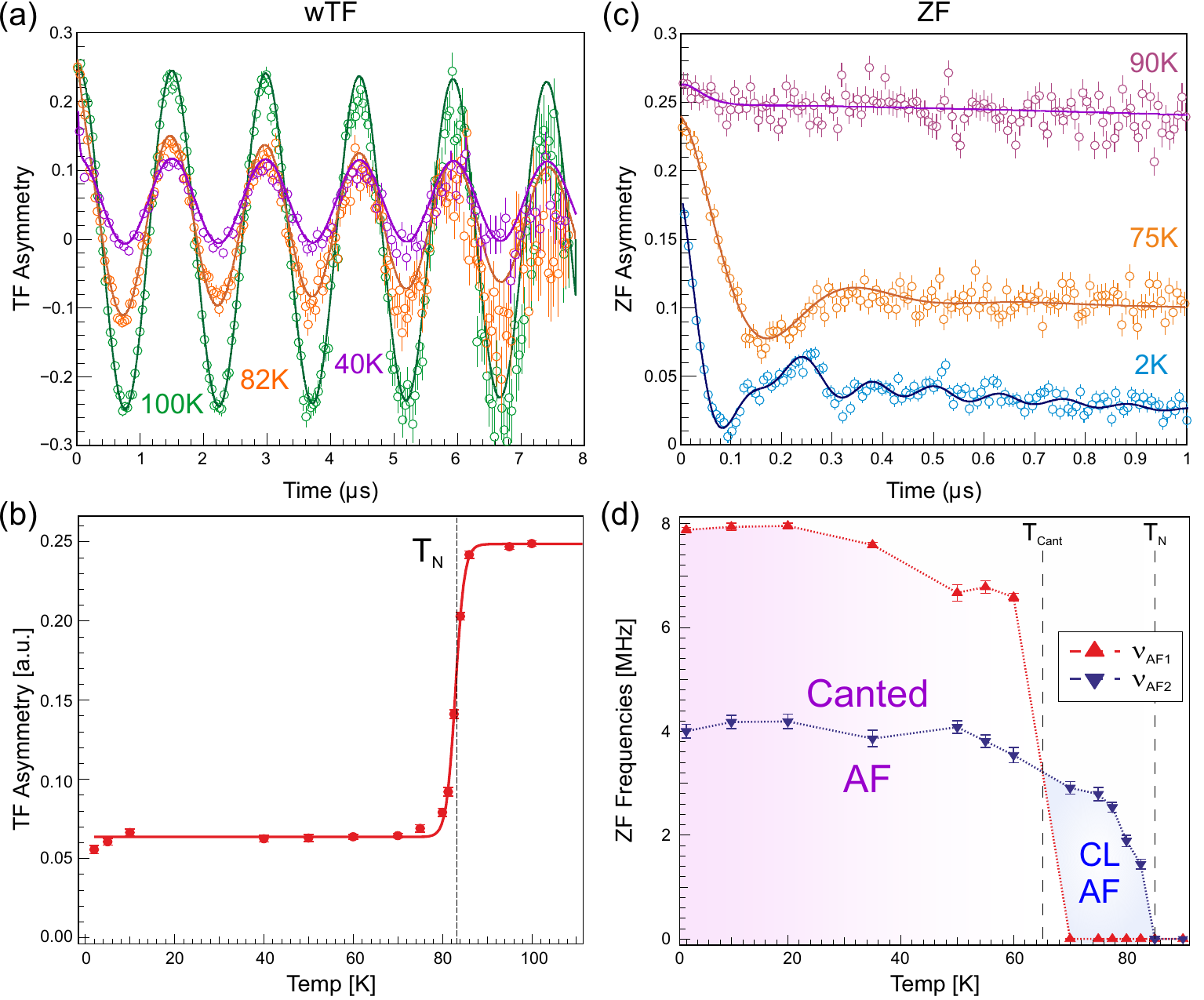}
  \end{center}
  \caption{(a) wTF muon time spectra at different temperatures. The black lines are fits to equation \ref{wtf}. (b) Temperature dependence of the transverse field asymmetry in  MgReO$_4$. The continuous line is a fit to the sigmoid function and the dashed line underlines the magnetic transition temperature. (c) ZF raw muon time spectra in the short time domain for different temperatures. The black lines are fits to equation \ref{zf1}. For clarity of display, the ZF spectra are shifted along the y axis. (d) Temperature dependence of the internal field frequencies. The dotted lines are guides to the eye and dashed vertical lines indicate the antiferromagnetic (AF) order transition ($T_{\rm N}\approx85$~K) into a collinear (CL) AF spin order, along with the spin reorientation temperature ($T_{\rm Cant}\approx65$~K) below which the spins arrange themselves in a proposed canted AF order.}
  \label{data}
\end{figure}

The weak transverse field (wTF) $\mu^+$SR time spectra at selected temperatures are shown in Figure \ref{data}(a). The wTF data is very well fitted [black solid lines in Figure \ref{data}(a)] by a muon polarization function described as follows:
\begin{eqnarray}
 A_0 \, P_{\rm TF}(t) &=& A_{\rm TF}\cos(2\pi \nu_{\rm TF}t + \frac{\pi \phi}{180})\cdot{}e^{(-\lambda_{\rm TF} t)}
\cr
 &+& A_{\rm tail}\cdot{}e^{(-\lambda_{\rm tail} t)}
\cr
 &+& A_{\rm AF}\cos(2\pi \nu_{\rm AF}t + \frac{\pi \phi_{AF}}{180})\cdot{}e^{(-\lambda_{\rm AF} t)}
\label{wtf}
\end{eqnarray}

The muon signal in the wTF regime consists of 3 terms: an oscillating signal with asymmetry A$_{\rm TF}$ reflecting the externally applied weak transverse field, a tail component, commonly found in powder samples (asymmetry A$_{\rm tail}$) with nearly zero relaxation rate ($\lambda_{\rm tail}$), and a fast relaxing term in the early time domain (AF component with asymmetry A$_{\rm AF}$), which becomes relevant in the temperature range of interest for the magnetic ordering. The phase for the AF component, $\phi_{AF}$ here shown for completeness, is fixed to zero. By plotting the transverse field asymmetry as a function of temperature [Fig. \ref{data} (b)], it is possible to obtain the magnetic transition temperature T$_N$ = (82.91 $\pm$ 0.07) K (consistent with T$_{C1}$ from susceptibility measurements) as the middle point of the sigmoid fitting function [continuous line in figure \ref{data}(b)].

The weak transverse field (wTF) oscillations in the signal are not completely suppressed at base temperature, i.e. the temperature dependence of the wTF-asymmetry does not reach the zero value at base temperature. This means that we have a small fraction of background signal that is not magnetically ordered even at lowest temperature. Most likely this background signal is a combination of muons stopping in the sample holder/envelope and/or a partial degradation of the sample. The latter could happen since the material is sensitive to air and moisture. Two of the powers of the $\mu^+$SR technique are its high sensitivity to low ordered magnetic moments along with its possibility to acquire data under true zero externally applied magnetic fields. Here we also acquired temperature dependent data using the so-called zero field (ZF) protocol. Selected ZF time spectra are presented in Fig.~\ref{data}(c), which reveal a clear oscillation at base-temperature ($T=2$~K). The ZF spectra were fitted with two exponentially damped cosine functions, both having a phase $\phi=0$:
\begin{eqnarray}
 A_0 \, P_{\rm ZF}(t) &=&
A_{\rm AF1}\cos(2\pi \nu_{\rm AF1}t + \frac{\pi \phi_{\rm AF1}}{180})\cdot{}e^{(-\lambda_{\rm AF1} t)}
\cr
 &+& A_{\rm AF2}\cos(2\pi \nu_{\rm AF2}t + \frac{\pi \phi_{\rm AF2}}{180})\cdot{}e^{(-\lambda_{\rm AF2} t)}
\cr
 &+& A_{\rm Fast}\cdot{}e^{(-\lambda_{\rm Fast} t)}
\cr
 &+& A_{\rm tail}\cdot{}e^{(-\lambda_{\rm tail} t)}.
\label{zf1}
\end{eqnarray}

Such function indicates the presence of a long range magnetic ordering with a magnetic unit cell commensurate to the symmetry of the crystal lattice. In addition, a fast relaxing exponential term, consistent with the presence of magnetic impurities (possibly metallic Re), and a tail component, usually found in powder samples, are also found in the ZF muon signal. The clear establishment of a static long range magnetic ordering clearly rules out the possibility of dimerization and formation of a spin singlet ground state in MgReO$_4$. The presence of two distinct frequencies $\nu_{AF1}$ and $\nu_{AF2}$ in the sample's internal field distribution usually denotes the presence of two nonequivalent crystallographic muon stopping sites. However, it is also possible that this is a result from one muon stopping site in combination with a complex magnetic structure (that could be either commensurate or incommensurate to the crystal lattice). With increasing temperature, it is found that there is a difference in the temperature dependencies of the two frequencies [see Fig.~\ref{data}(d)]. While $\nu_{AF1}$ is rather abruptly suppressed around 65 K, $\nu_{AF2}$ survives up until $T_{\rm N}\approx85$~K, displaying a clear second order-like phase transition. Since the wTF data don't show any additional magnetic transitions between 60 K and 70 K the possibility of a magnetic phase separation in the sample can be completely ruled out (determination of magnetic volume fractions is another power of the $\mu^+$SR technique). The change from one to two frequencies at 65 K could be the result of a structural transition resulting in a change in the muon stopping sites. However, from the $C_p$ data shown in Fig.~\ref{bulk}(b-d), there are no indications of a structural transition in this temperature range. Another, here more likely, scenario is that a spin reorientation occurs, creating a complex magnetic structure below 65 K. The zero phase of the ZF oscillations tells us that it should be a commensurate spin structure. Consequently, this suggests that a canted AF structure is formed below $T_{\rm Cant}\approx65$~K, while a collinear (CL) AF phase is present for temperatures $T_{\rm Cant}<T<T_{\rm N}$ [see Fig.~\ref{data}(d)].

\begin{figure}[h]
  \begin{center}
    \includegraphics[scale=0.77]{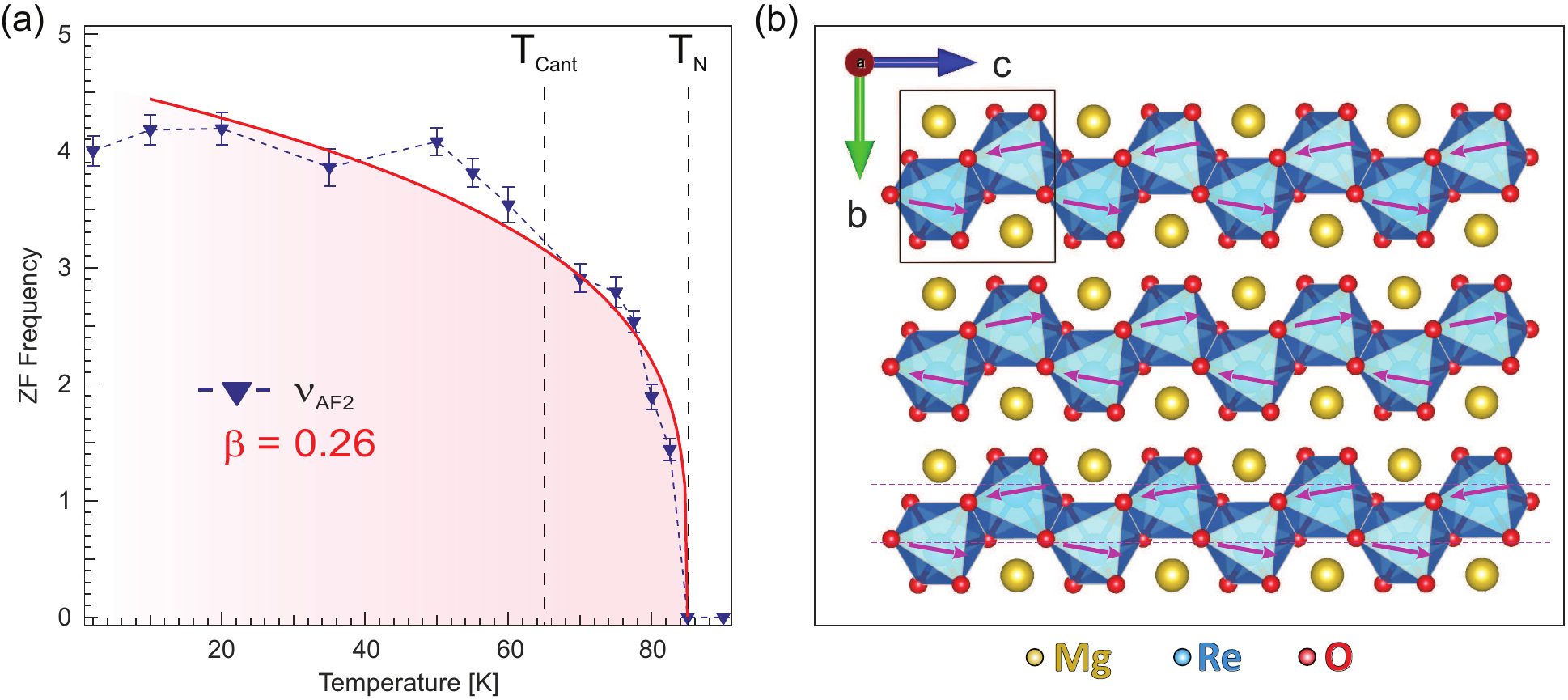}
  \end{center}
  \caption{(a) Order parameter of the magnetic phase transition represented by the slower oscillation frequency ($\nu_{\rm AF2}$). The red solid line is a fit to a mean-field theory power law: $\nu \propto (1-\frac{T}{T_N})^{\beta}$, which yields $\beta=0.26$. (b) Schematically presented canted spin structure (canting is exaggerated for display purposes) that is potentially established in MgReO$_4$ below $T_{\rm Cant}\approx65$~K. Dashed horizontal pink lines are guides to the eye to emphasize the canting of the spins away from the collinear spin structure.}
  \label{ord}
\end{figure}

The occurrence of canted AF orders is usually a result of competing magnetic interactions, in qualitative agreement with other systems showing a similar behavior \cite{blundell2008magnetism,NocerinoNaCrO,Sugiyama_2011,Sugiyama_2012,Thede_2014,Forslund_2020}. No detailed structural refinement of the atomic positions in MgReO$_4$ is reported. However, assuming a structure similar to the sister compound MnReO$_4$ \cite{bramnik2005preparation}, the Re spins in the zig-zag chain could be subjected to exchange interactions with nearest neighbor as well as next nearest neighbor spins. This is similar to the double tungstates analogues of MgReO$_4$ \cite{nyam2008magnetic}. The temperature dependence of the ZF-$\mu^+$SR frequencies reflects the order parameter of the magnetic phase transition in spin ordered materials \cite{yaouanc2011muon,Mansson_2012,Sugiyama_2012}. By fitting the second order-like muon frequency $\nu_{AF2}$ in MgReO$_4$, with a mean field theory power law $\nu \propto (1-\frac{T}{T_N})^{\beta}$, it is possible to obtain an indication about the nature of magnetic transition in this material [Fig.~\ref{ord}(a)]. The value of the critical exponent obtained from the fit, $\beta=0.26\pm0.03$, suggests that the magnetism in MgReO$_4$ cannot be categorized under any specific universality class, hereby endorsing the conjecture of a non-conventional magnetic nature of this material. It is worth mentioning that, if we consider the lower extreme of the relatively large uncertainty on the value of $\beta$, MgReO$_4$ could be regarded as an Heisenberg XY antiferromangnet. However, the mean value of the critical exponent $\beta=0.26$, is consistent with the aforementioned scenario of competing magnetic interactions. Therefore, the possibility of an unconventional critical behavior in MgReO$_4$ seems rather likely. Combining the presence of competing magnetic interactions and the itinerant nature of the Re$^{6+}$ 5$d^1$ electrons, leads to the conclusion that formation of a CL AF order below $T_{\rm N}\approx85$~K which is re-arrange into a canted AF spin order below $T_{\rm Cant}\approx65$~K is possible in MgReO$_4$. As already mentioned, this scenario is also coherent with spin orders that are commensurate to the crystal structure, which is expected from the zero phase in the cosine fit of the ZF data. A very crude schematic representation of one of the possible spin configurations in MgReO$_4$, compatibly with these observations, is displayed in Fig.~\ref{ord}(b).

Finally, we do not find any oscillations or clear magnetic signal above the AF transition temperature $T_{\rm C1}=T_{\rm N}\approx85$~K. Consequently, the anomaly observed in the susceptibility measurements [$\chi(T)$] as well as heat capacity [$C_p(T)$] at $T_{\rm C2}\approx280$~K (see Fig.~\ref{bulk}) is most likely not related to an additional magnetically ordered state. It is instead more likely related to a structural transition, which could easily be checked via, e.g. X-ray diffraction (XRD).

\section{Conclusion}
In this work we present the the first study of the magnetic properties of the wolframite insulator MgReO$_4$, investigated by both bulk methods as well as the $\mu^+$SR technique. From the detailed analysis of sensitive zero-field $\mu^+$SR data we find that a collinear antiferromagnetic (AF) order is first formed below $T_{\rm N}\approx85$~K. Such spin order is then transformed via a spin-reorientation into a canted AF phase below $T_{\rm Cant}\approx65$~K. Our study clearly reveal that such a scenario is very likely, considering the itinerant nature of the Re$^{6+}$ 5$d^1$ electrons and indications for the presence of competing magnetic interactions. Nevertheless, further experimental investigations (e.g. neutron diffraction), as well as modeling of muon stopping sites, would be needed to reveal the true magnetic ground state and detailed spin structure of this interesting material. Finally, computer modeling could be utilized for extracting the theoretical magnetic exchange interactions ($J/J'/...$) in order to also clarify the possible formation of multipolar magnetic ordering in MgReO$_4$.

\section{Acknowledgments}
The $\mu^+$SR measurements were performed at the instrument GPS of the Swiss Muon Source (S$\mu$S), at the Paul Scherrer Institute in Villigen, Switzerland. The authors wish to thank the staff of PSI for their valuable support. This research is funded by the Swedish Foundation for Strategic Research (SSF) within the Swedish national graduate school in neutron scattering (SwedNess), as well as the Swedish Research Council VR (Dnr. 2021-06157), and the Carl Tryggers Foundation for Scientific Research (CTS-18:272). JS was supported by the Japan Society for the Promotion Science (JSPS) KAKENHI Grant No.No. JP18H01863 and  JP20K21149. All images involving crystal structure were made with the VESTA software \cite{momma}, the parameter fitting has been performed with the software Igorpro \cite{igor} and the $\mu^+$SR data were fitted using the software package \textit{musrfit} \cite{musrfit}.

\section{References}
\providecommand{\newblock}{}


\begin{thebibliography}{10}
\expandafter\ifx\csname url\endcsname\relax
  \def\url#1{{\tt #1}}\fi
\expandafter\ifx\csname urlprefix\endcsname\relax\def\urlprefix{URL }\fi
\providecommand{\eprint}[2][]{\url{#2}}

\bibitem{Jana_2019}
Jana S, Aich P, Kumar P~A, Forslund O~K, Nocerino E, Pomjakushin V, M{\aa}nsson
  M, Sassa Y, Svedlindh P, Karis O, Siruguri V and Ray S 2019 {\em Scientific
  Reports\/} {\bf 9} \urlprefix\url{https://doi.org/10.1038/s41598-019-54427-0}

\bibitem{Konpap}
Papadopoulos K, Forslund O~K, Nocerino E, Johansson F~O~L, Simutis G, Matsubara
  N, Morris G, Hitti B, Arseneau D, Orain J~C, Pomjakushin V, Svedlindh P,
  Andreica D, B\"{o}rjesson L, Sugiyama J, Månsson M and Sassa Y 2021
  \urlprefix\url{https://arxiv.org/abs/2111.05920}

\bibitem{Forslund_2020}
Forslund O~K, Papadopoulos K, Nocerino E, Morris G, Hitti B, Arseneau D,
  Pomjakushin V, Matsubara N, Orain J~C, Svedlindh P, Andreica D, Jana S,
  Sugiyama J, M{\aa}nsson M and Sassa Y 2020 {\em Physical Review B\/} {\bf
  102} \urlprefix\url{https://doi.org/10.1103/physrevb.102.144409}

\bibitem{sleight1975new}
Sleight A 1975 {\em Inorganic Chemistry\/} {\bf 14} 597--599

\bibitem{baur1992coreo4}
Baur W~H, Joswig W, Pieper G and Kassner D 1992 {\em Journal of Solid State
  Chemistry\/} {\bf 99} 207--211

\bibitem{urushihara2021structural}
Urushihara D, Asaka T, Fukuda K, Nakayama M, Nakahira Y, Moriyoshi C, Kuroiwa
  Y, Forslund O~K, Matsubara N, Mansson M {\em et~al.\/} 2021 {\em Inorganic
  Chemistry\/} {\bf 60} 507--514

\bibitem{chen2010exotic}
Chen G, Pereira R and Balents L 2010 {\em Physical Review B\/} {\bf 82} 174440

\bibitem{bramnik2001preparation}
Bramnik K, Ehrenberg H and Fuess H 2001 {\em Journal of Solid State
  Chemistry\/} {\bf 160} 45--49

\bibitem{bramnik2003preparation}
Bramnik K~G, Ehrenberg H, Dehn J~K and Fuess H 2003 {\em Solid state
  sciences\/} {\bf 5} 235--241

\bibitem{bramnik2005preparation}
Bramnik K, Ehrenberg H, Buhre S and Fuess H 2005 {\em Acta Crystallographica
  Section B: Structural Science\/} {\bf 61} 246--249

\bibitem{patent}
Sleight A~W 1977 United states patent 4027004

\bibitem{popov2003large}
Popov G, Greenblatt M and Croft M 2003 {\em Physical Review B\/} {\bf 67}
  024406

\bibitem{blundell1999spin}
Blundell S 1999 {\em Contemporary Physics\/} {\bf 40} 175--192

\bibitem{Blundell_2021}
Blundell S~J, De~Renzi R, Lancaster T and Pratt F~L (eds) 2021 {\em Muon
  Spectroscopy\/} (London, England: Oxford University Press)

\bibitem{blundell2008magnetism}
Blundell S, Lancaster T, Baker P, Hayes W, Pratt F, Atake T, Rana D~S and Malik
  S 2008 {\em Physical Review B\/} {\bf 77} 094424

\bibitem{NocerinoNaCrO}
Nocerino E {\em et~al.\/} 2022 \textit{Revised Magnetic structure and
  tricritical behavior of the CMR Compound NaCr$_2$O$_4$ investigated with High
  Resolution Neutron Diffraction and $\mu^+$SR}

\bibitem{Sugiyama_2011}
Sugiyama J, M{\aa}nsson M, Ofer O, Kamazawa K, Harada M, Andreica D, Amato A,
  Brewer J~H, Ansaldo E~J, Ohta H, Michioka C and Yoshimura K 2011 {\em
  Physical Review B\/} {\bf 84}
  \urlprefix\url{https://doi.org/10.1103/physrevb.84.184421}

\bibitem{Sugiyama_2012}
Sugiyama J, Nozaki H, M{\aa}nsson M, Pr{\v{s}}a K, Amato A, Isobe M and Ueda Y
  2012 {\em Physics Procedia\/} {\bf 30} 186--189
  \urlprefix\url{https://doi.org/10.1016/j.phpro.2012.04.069}

\bibitem{Thede_2014}
Thede M, Mannig A, M{\aa}nsson M, H\"{u}vonen D, Khasanov R, Morenzoni E and
  Zheludev A 2014 {\em Physical Review Letters\/} {\bf 112}
  \urlprefix\url{https://doi.org/10.1103/physrevlett.112.087204}

\bibitem{nyam2008magnetic}
Nyam-Ochir L, Ehrenberg H, Buchsteiner A, Senyshyn A, Fuess H and Sangaa D 2008
  {\em Journal of magnetism and magnetic materials\/} {\bf 320} 3251--3255

\bibitem{yaouanc2011muon}
Yaouanc A and De~Reotier P~D 2011 {\em Muon spin rotation, relaxation, and
  resonance: applications to condensed matter\/} 147 (Oxford University Press)

\bibitem{Mansson_2012}
M{\aa}nsson M, Pr{\v{s}}a K, Sugiyama J, Andreica D, Luetkens H and Berger H
  2012 {\em Physics Procedia\/} {\bf 30} 142--145
  \urlprefix\url{https://doi.org/10.1016/j.phpro.2012.04.059}

\bibitem{momma}
Momma K and Izumi F 2008 {\em Journal of Applied Crystallography\/} {\bf 41}
  653--658

\bibitem{igor}
 2017 (\textit{Preprint}
  \eprint{http://www.wavemetrics.com/products/igorpro/igorpro.htm})

\bibitem{musrfit}
{A Suter and B~M Wojek} 2012 {\em Phys. Proc.\/} {\bf 30} 69
  \urlprefix\url{http://www.dx.doi.org/10.1016/j.phpro.2012.04.042}

\end{thebibliography}
\end{document}